\documentclass[twocolumn,aps,]{openjournal}
\usepackage{enumitem}
\usepackage{graphicx}
\usepackage{bm}
\usepackage[colorlinks=True]{hyperref}
\usepackage{natbib}
\usepackage{booktabs}
\usepackage{amsmath}
\usepackage[mathlines]{lineno}
\begin{document}

\title{Minkowski Functionals in Joint Galaxy Clustering \& Weak Lensing Analyses}

\author{Nisha Grewal$^1 \ast$, Joe Zuntz$^1$, Tilman Tr\"oster$^1$, and Alexandra Amon$^2$}
\affiliation{
    $^1$Institute for Astronomy, University of Edinburgh, Royal Observatory, Blackford Hill, Edinburgh, EH9 3HJ, UK\\
    $^2$Institute of Astronomy, University of Cambridge, Madingley Road, Cambridge, CB3 0HA, UK
}

\email{$\ast$ nisha.grewal@ed.ac.uk}

\begin{abstract}
         We investigate the inclusion of clustering maps in a weak lensing Minkowski functional (MF) analysis of DES-like and LSST-like simulations to constrain cosmological parameters. The standard 3x2pt approach to lensing and clustering data uses two-point correlations as its primary statistic; MFs, morphological statistics describing the shape of matter fields, provide additional information for non-Gaussian fields. 
         Previous analyses have studied MFs of lensing convergence maps; in this project we explore their simultaneous application to clustering maps. We employ a simplified linear galaxy bias model, and using a lognormal curved sky measurement and Monte Carlo Markov Chain (MCMC) sampling process for parameter inference, we find that MFs do not yield any information in the $\Omega_{\rm m}$ -- $\sigma_8$ plane not already generated by a 3x2pt analysis. However, we expect that MFs should improve constraining power when nonlinear baryonic and other small-scale effects are taken into account. As with a 3x2pt analysis, we find a significant improvement to constraints when adding clustering data to MF-only and MF$+C_\ell$ shear measurements, and strongly recommend future higher order statistics be measured from both convergence and clustering maps.

\end{abstract}

\maketitle

\section{\label{sec:level1}Introduction}

Statistics measured from gravitational lensing have been effective in probing the growth of large-scale structure and the behaviour of gravity on cosmological scales. Ongoing surveys like the Dark Energy Survey\footnote{https://www.darkenergysurvey.org/} \citep[DES; ][]{des_survey}, Hyper Supreme-Cam\footnote{https://www.naoj.org/Projects/HSC/} \citep[HSC; ][]{hsc_survey}  survey, and Kilo-Degree Survey\footnote{http://kids.strw.leidenuniv.nl/} \citep[KiDS; ][]{kids_survey} observe tens to hundreds of millions of galaxies. Covering 5000 deg$^2$, 1400 deg$^2$, and 1500 deg$^2$ respectively, these surveys construct detailed galaxy catalogs \citep{des_cat,kids_cat,hsc_cat} and high resolution weak lensing maps \citep[e.g.][]{des_map} and have extracted significant cosmological information from their data \citep{kids_cosmo,hsc_cosmo,des_cosmo}.

Cosmic shear is independent of galaxy bias and sensitive to the geometry and evolution of the Universe, making it particularly informative in the study of large scale structure \citep{hsc_shear2,hsc_shear,kids_shear2,kids_shear,des_shear}. Using the $\Lambda$CDM model, which takes the assumption that the Universe comprises baryonic matter, cold dark matter, and dark energy, two-point statistics of weak lensing and galaxy clustering have been used successfully to constrain cosmological parameters that describe the model. 

Measuring both convergence and clustering information using two-point statistics, such as correlation functions or the angular power spectra $C_{\ell}$, has proven particularly effective, yielding high precision constraints on cosmological parameters $\Omega_{\rm m}$ and $\sigma_8$ \citep{kids_cosmo,hsc_cosmo,des_cosmo}.

The two-point measurements described above encode all the information present in purely Gaussian random fields. In the non-Gaussian fields generated by nonlinear structure formation, however, other (higher-order) statistics can contain additional information. One example is peak statistics, which, when applied to weak lensing maps, can break degeneracies between cosmological parameters while remaining robust against systematic effects \citep[e.g.][]{dietrich_2010,kratochvil_2010,liu_2015,Kacprzak_2016,KIDS_peak,martinet_2018,peel_2018,ajani_2020,zurcher2021}. Other higher order statistics of galaxy density, galaxy shear, and cosmic microwave background (CMB) lensing like three point correlation functions and bi-spectra \citep[e.g.][]{3ptcf_2003,VAFAEI,3ptcf_2011,Petri,3ptcf_2014,secco3pt} as well as moments \citep[e.g.][]{moment1,moment2,moment3,moment4,moment6,moment5,moment7,gatti_moments_2021} are able to achieve even higher precision.
 
Many statistics, including these, are measured from maps generated from galaxy catalogues.  Convergence maps, which are projected measures of mass along the line of sight, are derived from shear catalogues of source galaxies. Clustering maps, which show the number density of galaxies, are built from lens galaxy position catalogues and are complementary tracers of cosmic structure \citep{clustering}.  Clustering maps have a higher signal to noise ratio, enabling more detailed results \citep{des_map}. However, clustering maps are only able to probe the distribution of galaxies, so galaxy bias must be modelled to relate this to the density of dark matter, which dominates the overall matter density \citep{des_cosmo}. 

The morphological descriptors \textit{Minkowski functionals} (MFs) are another higher-order statistic applicable in this context \citep{minkowski}. MFs are unbiased functional integrals over fields, have low variance, and have been applied previously to non-Gaussian convergence maps \citep{first_mf,kratochvil,Petri,deep_minkowski}, CMB data \citep[e.g.][and references thereto]{cmb_mf}, and 3D density fields from spectroscopic data \citep{3Dclust_1,3Dclust_2,3Dclust_3,3Dclust_4}.  For a purely Gaussian field within a $\Lambda$CDM model, MFs and $C_{\ell}$ contain the same information for $\Omega_m$ and $\sigma_8$. Adding MFs to a single lensing field analysis has been shown to greatly improve constraints on cosmological parameters, and their angular-scale dependence is particularly effective in separating primordial non-Gaussianities from gravity-induced non-Gaussianities \citep{munshi,primordial_mf,vicinanza}.  Because they are sensitive to small-scale structure, MFs have strong constraining power. They have the potential to have additional resilience to some systematic uncertainties \citep{non-Gaus}.  Multiple MFs at different smoothing scales probe more information than a single set of MFs, and more analytical treatments like Hermite polynomials \citep{3Dclust_4} could extract more information as well.

In this paper, we investigate the effectiveness of the application of the first three MFs to multi-field simulations of both convergence and clustering maps, with a focus on forecasting potential constraining power using upcoming datasets. Future surveys will be even wider and deeper than current surveys, have the capabilities to build more informed weak lensing maps, and provide even more cosmological information. The Legacy Survey Space and Time (LSST) at the Vera C. Rubin Observatory\footnote{https://www.lsst.org/} \citep{lsst_book,lsst_desc} and the space telescopes Euclid\footnote{https://www.euclid-ec.org/} \citep{euclid} and Nancy Grace Roman\footnote{https://roman.gsfc.nasa.gov/} \citep{roman} will map nearly half the sky, and so enable dramatically strong constraints on the dark sector.  The statistical approach we develop here can eventually be applied to such datasets.

In Section \ref{section:Formalism} we cover the theory of MFs and power spectra $C_\ell$, in Section \ref{section:Methodology} we go over the process by which we use an MCMC to generate cosmological parameter posterior values, in Section \ref{section:Results} we show the constraints on cosmological parameters from various scenarios, and we conclude in Section \ref{section:Conclusion}.

\section{Formalism/Theory}
\label{section:Formalism}

\subsection{Minkowski Functionals}

Minkowski functionals are mathematical descriptors of the topology of continuous fields \citep{minkowski,non-Gaus}. For 2D random fields, there are three functionals, quantifying area, perimeter, and mean curvature of an excursion set \citep[the region of a field above a given threshold;][]{deep_minkowski}.  The first three MFs are defined as
\begin{equation}
    V_0(\nu) = \frac{1}{A} \int\limits_A \Theta(\alpha(\textbf{x})-\nu)d\phi d\theta ,
    \label{v0}
\end{equation}

\begin{equation}
    V_1(\nu) = \frac{1}{4A} \int\limits_A \delta(\alpha(\textbf{x})-\nu)\sqrt{\alpha_\phi^2 + \alpha_\theta^2} d\phi d\theta ,
    \label{v1}
\end{equation}

\begin{multline}
    V_2(\nu) = \frac{1}{2\pi A} \int\limits_A \delta(\alpha(\textbf{x})-\nu) \\
    \left( \frac{2\alpha_\phi \alpha_\theta \alpha_{\phi\theta}-\alpha_\phi^2 \alpha_{\theta\theta}-\alpha_\theta^2 \alpha_{\phi\phi}}{\alpha_\phi^2 + \alpha_\theta^2} \right) d\phi d\theta ,
    \label{v2}
\end{multline}
where \textbf{x} is the location in the field, $\nu$ is a chosen threshold, $A$ is the total area of the map, $\phi$ and $\theta$ are polar coordinates, $\alpha(\textbf{x})$ is the field value in two dimensions, and $\alpha_\phi$, $\alpha_\theta$, $\alpha_{\phi\theta}$, $\alpha_{\theta\theta}$, and $\alpha_{\phi\phi}$ are derivatives of the field \citep{minkowski,Petri}. In Eq. \eqref{v0} the Heaviside function $\Theta$ selects the field region above the threshold, whose area is calculated;  Eq. \eqref{v1} employs the Dirac delta function to select the perimeter of that region whose heights are at the same level as the threshold; similarly, Eq. \eqref{v2} finds the curvature of the boundary, which also describes the connectivity of the field \citep{minkowski}. Since the fields we use are not normalised to have unit variance, the sensitivity of $\sigma_8$ comes from the overall amplitudes of $V_1$ and $V_2$.

\begin{figure*}[ht]
    \centering
    \includegraphics[width=\textwidth]{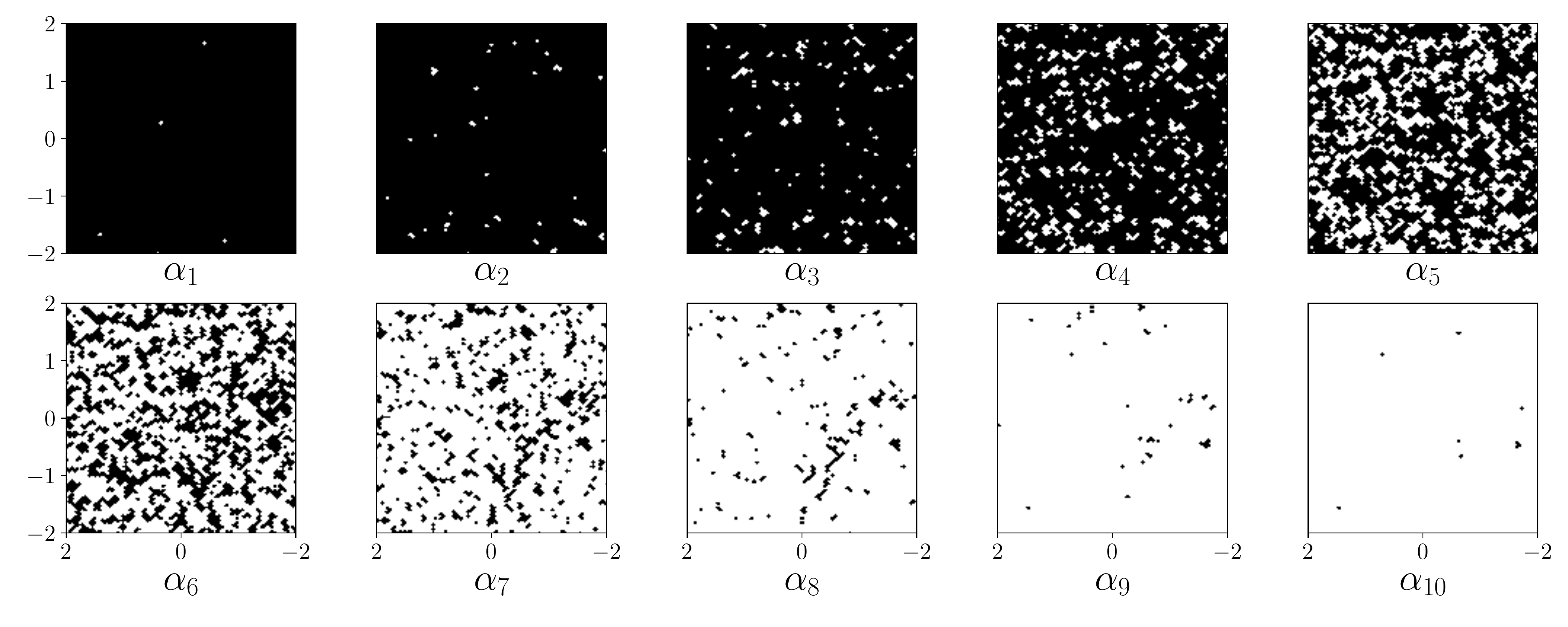}
    \caption{Excursion sets at 10 threshold values $\alpha$ of a simulated convergence map, where the regions under the threshold are marked in white and above in black.}
    \label{fig:e_sets}
\end{figure*}

\begin{figure*}[ht]
    \centering
    \includegraphics[width=\textwidth]{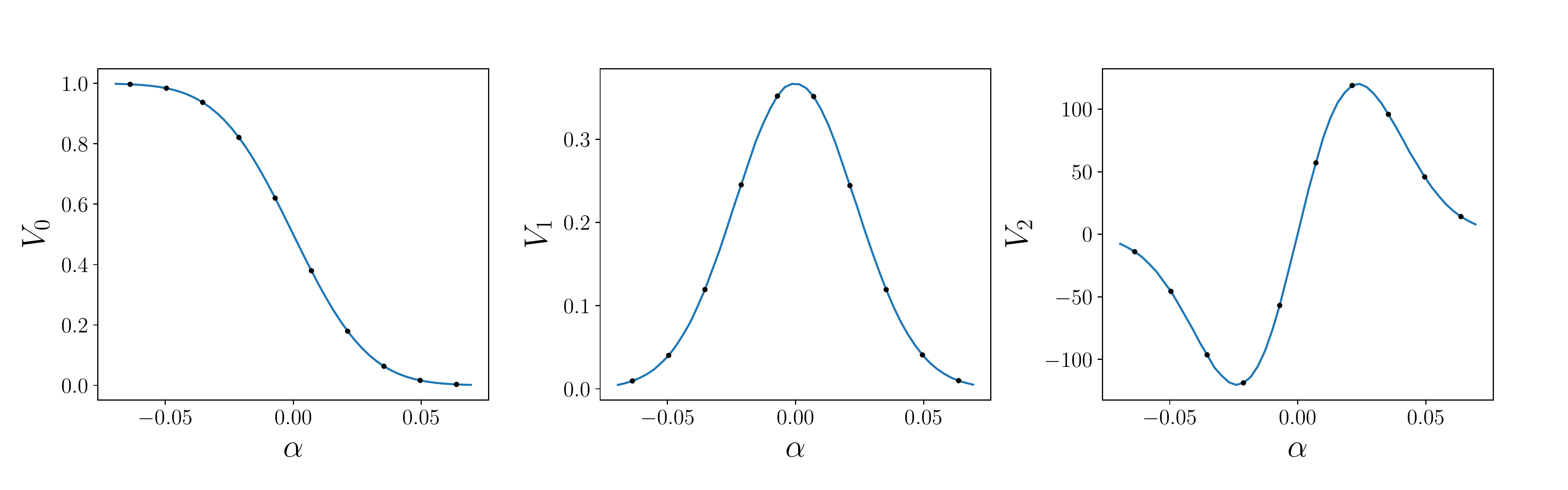}
    \caption{Sample of the first three MFs $V_0$, $V_1$, and $V_2$ calculated from a simulated weak lensing convergence map using fiducial $\Omega_{\rm m}$ and $\sigma_8$ values. Each black dot on the curve corresponds to an excursion set in Figure \ref{fig:e_sets}, and the three MFs on the $y$-axis describe the perimeter, area, and curvature of the mass distribution \citep{Petri}.}
    \label{fig:mf}
\end{figure*}

Figure  \ref{fig:e_sets} and \ref{fig:mf} illustrate the three MFs.  The excursion sets in Figure \ref{fig:e_sets} give rise to the three MFs plotted as a function of the threshold value in \ref{fig:mf}.  If we visualize the original 2D map as a topographic height map, the excursion sets are the regions above a given altitude.

\subsection{Power Spectra}

We will compare the constraining power of MFs to the standard 2D power spectrum $C_{\ell}$. The angular power spectrum $C_{\ell}^{xy}$ measures the scale-dependent structure of two fields $x, y$, which in this case are either weak lensing convergence or galaxy density. Assuming an isotropic system, 
\begin{equation}
     \langle a_{lm}^{x}, a_{l'm'}^{y*} \rangle = \delta_{\ell\ell'}\delta_{mm'}C_\ell^{xy}, 
\end{equation}
where $a_{lm}$ is the spherical harmonic transform of a field \citep{angularpowerspectra}. We project a 3D field $X(\vec{\chi})$ via a weighting function $w_{x}(\chi)$ into the 2D field $x$ \citep{weak_matt}:
\begin{equation}
    x(\vec{\theta}) = \int_{0}^{\chi_{\rm S}} d\chi w_{x}(\chi) X(\vec{\theta},\chi).
\end{equation}

Using Limber's approximation in a flat Universe, we can convert $P_{XY}$, the 3D (cross-)power spectrum of $X$,$Y$, into the 2D angular power spectrum of $x$,$y$:
\begin{equation}
    C_{\ell}^{xy} = \int_{0}^{\chi_{\rm S}} d\chi \frac{w_x(\chi) w_y(\chi)}{\chi^2}P_{XY} \left(\frac{\ell + \frac{1}{2}}{\chi}, z(\chi) \right),
\end{equation}
where $\ell$ is the scalar multipole, $\chi_{\rm S}$ is the comoving distance to the source plane, $\chi$ is the comoving angular diameter distance, and $w$ is a radial kernel function \citep{Limber,weaklens,weak_matt,des_cosmo}. The kernel function for convergence is given by
\begin{equation}
    w_\kappa^j(\chi) = \frac{3\Omega_{\mathrm{m}}H_0^2}{2} \int_\chi^{\chi_H} d\chi' n_s^j(\chi') \frac{\chi}{a(\chi)} \frac{\chi'-\chi}{\chi'},
\end{equation}
where $\Omega_{\mathrm{m}}$ is the present-day matter density parameter, $H_0$ is the Hubble parameter, $\chi_H$ is the comoving distance to the horizon, $n(\chi)$ is the source galaxy number density distribution, and $a(\chi)$ is the scale factor \citep{weaklens}. For clustering, the kernel function is simpler:
\begin{equation}
    n(\chi) = n(z(\chi))\frac{dz}{d\chi}
\end{equation}
\citep{clustering}. We do not use cross-correlations of maps in our simulations (see Section \ref{sssec:redshift}), so the statistical structure of our power spectrum analysis does not fully replicate 3x2pt analysis.

For Gaussian random fields, such as those in the nearly-homogeneous early Universe, $C_{\ell}$ contain all the information present \citep{Coil_2013}. While they are still useful for more general non-Gaussian late-time fields, they do not fully describe the statistics of such systems.

\section{Methodology}
\label{section:Methodology}

In this paper we introduce the application of MFs to clustering maps. We compare the constraints on cosmological parameters from convergence maps, clustering maps, and a combination of the two. To measure these constraints, we  explore the space of cosmological parameters with an MCMC process.

\subsection{Simulations} 
In this work we generate curved sky lognormal simulations, which let us compute exact derivatives of the field. The lognormal sky simulations are relatively quick to evaluate and provide some degree of non-Gaussian signal. Our analysis has the flexibility to be measured from more sophisticated nonlinear, non-Gaussian simulations in the future. 

\subsubsection{Redshift}
\label{sssec:redshift}

\begin{figure*}[ht]
\centering
\begin{minipage}{\columnwidth}
    \includegraphics[width=\columnwidth]{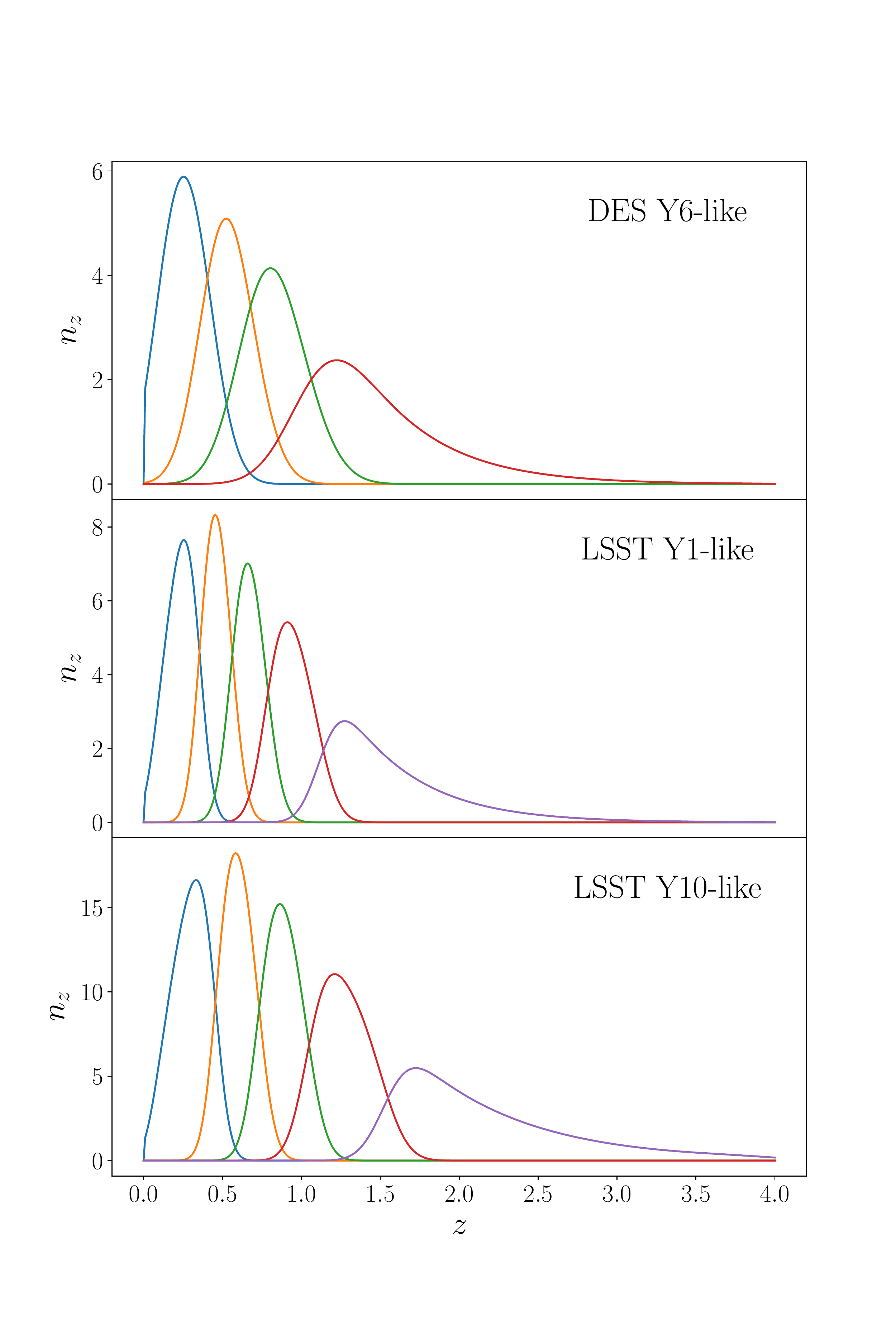}
\end{minipage}
\hfill
\begin{minipage}{\columnwidth}
    \includegraphics[width=\columnwidth]{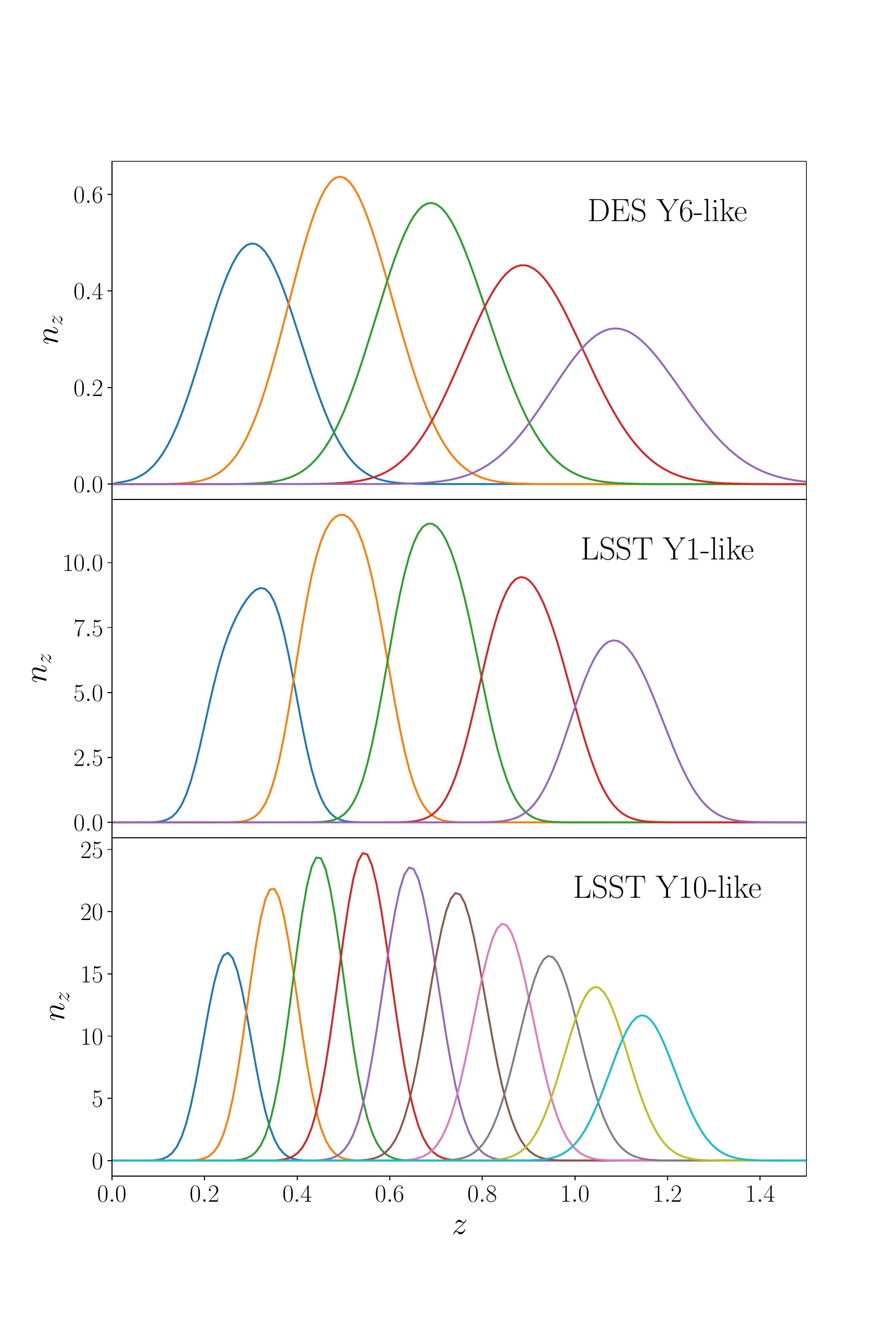}
\end{minipage}
\hfill
\caption{$n_z$ redshift bins for DES Y6-like, LSST Y1-like, and LSST Y10-like \citep{des_redshifts,n_zhang}. Convergence distributions are displayed on the left, and clustering distributions are on the right.}
\label{fig:redshift}
\end{figure*}

To build these simulated maps we first start from DES Y6-like, LSST Y1-like, and LSST Y10-like tomographic number densities based on those in \citet{n_zhang}, which are shown in Figure \ref{fig:redshift} and Table \ref{tab:redshifts}. These $n(z)$ values are generated from an underlying true redshift distribution of the form $n_\mathrm{true}(z) \propto z^2 \exp{\left(z/z_0\right)^a}$, divided into equal density tomographic bins and then convolved with a Gaussian error distribution with width $\sigma(z) = (1 + z) \sigma_z$. Different values of the parameters $a$ and $\sigma_z$ and the number of bins and total density are used for our three different scenarios, DES Y6-like, LSST Y1-like, and LSST Y10-like, using the values given in Table 1 of \citet{n_zhang}.

\begin{table*}[t]
    \centering
    \begin{tabular}{p{1.9cm}p{1.25cm}p{1.25cm}p{1.25cm}p{1.25cm}p{1.25cm}p{1.25cm}p{1.25cm}p{1.25cm}p{1.25cm}p{1.25cm}}
    \toprule
        &\multicolumn{1}{c}{Bin 1}&\multicolumn{1}{c}{Bin 2}&\multicolumn{1}{c}{Bin 3}&\multicolumn{1}{c}{Bin 4}&\multicolumn{1}{c}{Bin 5}&\multicolumn{1}{c}{Bin 6}&\multicolumn{1}{c}{Bin 7}&\multicolumn{1}{c}{Bin 8}&\multicolumn{1}{c}{Bin 9}&\multicolumn{1}{c}{Bin 10}\\
        \midrule 
        DES Y6-like & \multicolumn{1}{c}{0.20-0.43} & \multicolumn{1}{c}{0.43-0.63} & \multicolumn{1}{c}{0.63-0.90} & \multicolumn{1}{c}{0.90-1.30}\\
        LSST Y1-like & \multicolumn{1}{c}{0.20-0.43} & \multicolumn{1}{c}{0.43-0.63} & \multicolumn{1}{c}{0.63-0.90} & \multicolumn{1}{c}{0.90-1.30}\\
        LSST Y10-like & \multicolumn{1}{c}{0.20-0.43} & \multicolumn{1}{c}{0.43-0.63} & \multicolumn{1}{c}{0.63-0.90} & \multicolumn{1}{c}{0.90-1.30}\\
        \midrule
        DES Y6-like&\multicolumn{1}{c}{1.17}&\multicolumn{1}{c}{1.28}&\multicolumn{1}{c}{1.41}&\multicolumn{1}{c}{1.55}&\multicolumn{1}{c}{1.69}& & & & & \\
        LSST Y1-like&\multicolumn{1}{c}{1.17}&\multicolumn{1}{c}{1.28}&\multicolumn{1}{c}{1.41}&\multicolumn{1}{c}{1.55}&\multicolumn{1}{c}{1.69} & & & & &\\
        LSST Y10-like &\multicolumn{1}{c}{1.13}&\multicolumn{1}{c}{1.19}&\multicolumn{1}{c}{1.25}&\multicolumn{1}{c}{1.32}&\multicolumn{1}{c}{1.38}&\multicolumn{1}{c}{1.45}&\multicolumn{1}{c}{1.51}&\multicolumn{1}{c}{1.58}&\multicolumn{1}{c}{1.65}&\multicolumn{1}{c}{1.72}\\
        \bottomrule
    \end{tabular}
    \caption{\textup{Convergence map redshift bins in the top three rows and clustering map galaxy bias values in the bottom three rows \citep{des_redshifts,n_zhang}.}}
    \label{tab:redshifts}
\end{table*}

\subsubsection{Map Simulation}
\label{sssec:flask}
Next we use the Dark Energy Science Collaboration's \citep[DESC;][]{desc} Core Cosmology Library \citep[\texttt{CCL};][]{ccl} to predict theory values for the convergence and clustering power spectra $C_\ell$. \texttt{CCL} takes the redshift distributions $n(z)$ and cosmological parameters and computes $C_\ell$, which we pass with the $n(z)$ to the Full-sky Lognormal Astro-fields Simulation Kit \citep[\texttt{FLASK};][]{flask} to simulate the convergence and clustering maps. \texttt{FLASK} generates continuous lognormal fields using spherical geometry \citep{flask}.

\texttt{FLASK} generates noisy clustering maps directly using the noise level we supply, and we manually add noise to the convergence maps it generates. In both cases we used Gaussian noise levels corresponding to the number densities in our scenarios. FLASK's mock fields, as with real fields, are more non-Gaussian for the clustering than for the lensing, since they are integrated over a narrower redshift kernel, and the code has specific corrections to allow a sensible joint analysis in this intermediate non-linear and non-Gaussian regime \citep{flask}.

We originally used more realistic Poisson noise in our clustering maps, but this caused problems when trying to fix the random seed (see below) in our simulations. Changing the power spectra that \texttt{FLASK} takes as input changes the cosmic variance and noise generated in the system. But while with Gaussian noise the same number of random variates must be generated even as the mean levels change, this is not the case for Poisson noise. This means that for Poisson noise, a small change to input power spectra no longer creates a small change to the resulting maps, since the sequence of random noise values changes significantly\footnote{An alternative solution would be to fix the mean of the noise field even as the signal changes.}.

\subsubsection{Simulation Inputs}
\label{sssec:sim_inputs}

For this project, we use DES Y6-like, LSST Y1-like, and LSST Y10-like scenarios to find constraints on cosmological parameters.  The cosmological parameter inputs we vary in the \texttt{FLASK} simulation are $\Omega_{\rm m}$ (total matter), $\sigma_8$ (amplitude of the matter power spectrum), $\Omega_{\rm b}$ (baryonic matter), $H_0$ (present-day rate of expansion of the Universe), and $n_s$ (spectral index). We also vary linear galaxy bias values for the different lens bins. The fiducial values can be seen in Table \ref{tab:fiducial}.

\begin{table}[ht]
    \centering
    \begin{tabular}{cccc} 
    \toprule
    Parameter & Fiducial & Min & Max \\
    \midrule
        $\Omega_{\rm m}$ & 0.3 & 0.1 & 0.6\\ 
        $\sigma_8$ & 0.8 & 0.3 & 1.2\\
        $\Omega_b$ & 0.048 & 0.047 & 0.049\\
        $H_0$ & 0.7 & 0.5 & 0.9\\
        $n_s$ & 0.96 & 0.9 & 1.1 \\
    \bottomrule
    \end{tabular}
    \caption{\textup{Fiducial values of cosmological parameters with the minimum and maximum values given to the MCMC generator.}}
    \label{tab:fiducial}
\end{table}

To test the constraining power of adding clustering maps to the model, we use three sets of tomographic maps: convergence maps, clustering maps, and a combination of the two. Generating these simulations is slow, so we optimised a model that was informative while (relatively) efficient.  We use $N_\mathrm{side} = 1024$ for the resolution parameter throughout, and follow \citet{Petri}'s choices for Gaussian smoothing $\theta_G$ and MF threshold count $N_\textrm{bins}$, which can be seen in Table \ref{tab:smoothing}. While smoothing of 1 arcminute may be the most informative level, this is computationally taxing and essentially unsmoothed in our fields, thus subject to more noise. Going to smaller scales comes with other challenges, such as modelling uncertainties due to nonlinear galaxy bias and baryon feedback. Instead we use 5 arcminutes as the fiducial value, which is still informative at the $\mathrm{N_{side}}$ = 1024 level. 

\begin{table}[]
    \centering
    \begin{tabular}{cccccccc}
        \toprule
        \multicolumn{7}{c}{Statistic Type} \\
        \midrule
        $\theta_G$ & $\mathrm{N_{side}}$ & $\mathrm{N_{bins}}$ & mask & $n_z$ bins & MF$/C_{\ell}$ & Map Type \\
        \midrule
        5 & 1024 & 10 & 0.125 & DES Y6-like& $C_{\ell}$ & Joint \\
        5 &1024 & 10 & 0.44 & LSST Y1-like &MF & Joint \\
        5 &1024 & 10 & 1 & LSST Y10-like & $C_{\ell}$+MF & Joint \\  
        \midrule
        \multicolumn{7}{c}{Map Type} \\
        \midrule
        $\theta_G$ & $\mathrm{N_{side}}$ & $\mathrm{N_{bins}}$ & mask & $n_z$ bins & MF$/C_{\ell}$& lens$/$clust\\
        \midrule
        5 & 1024 & 10 & 0.125 & DES Y6-like& Joint & Lensing \\
        5 &1024 & 10 & 0.44 & LSST Y1-like& Joint & Clustering \\
        5 &1024 & 10 & 1 & LSST Y10-like & Joint & Joint \\  
        \midrule
        \multicolumn{7}{c}{Smoothing} \\
        \midrule
        $\theta_G$ & $\mathrm{N_{side}}$ & $\mathrm{N_{bins}}$ & mask & $n_z$ bins & MF$/C_{\ell}$& lens$/$clust\\
        \midrule
        1 & 1024 & 10 & 0.44 & LSST Y1-like& Joint & Joint \\
        5 & 1024 & 10 & 0.44 & LSST Y1-like& Joint & Joint \\
        15 & 1024 & 10 & 0.44 & LSST Y1-like& Joint & Joint \\
        \toprule
        \multicolumn{7}{c}{Survey} \\
        \midrule
        $\theta_G$ & $\mathrm{N_{side}}$ & $\mathrm{N_{bins}}$ & mask & $n_z$ bins & MF$/C_{\ell}$ & lens$/$clust\\
        \midrule
        5 & 1024 & 10 & 0.125 & DES Y6-like & Joint & Joint \\
        5 &1024 & 10 & 0.44 & LSST Y1-like& Joint & Joint \\
        5 &1024 & 10 & 0.44 & LSST Y10-like & Joint & Joint \\        
        \bottomrule
    \end{tabular}
    \caption{\textup{Scenarios considered in this paper: comparing statistic type, map type, Gaussian smoothing $\theta_G$ at 1, 5, and 15 arcminutes; and survey model: DES Y6-like, LSST Y1-like, and LSST Y10-like.}}
    \label{tab:smoothing}
\end{table}

\begin{figure}[ht]
    \centering
    \includegraphics[width=\columnwidth]{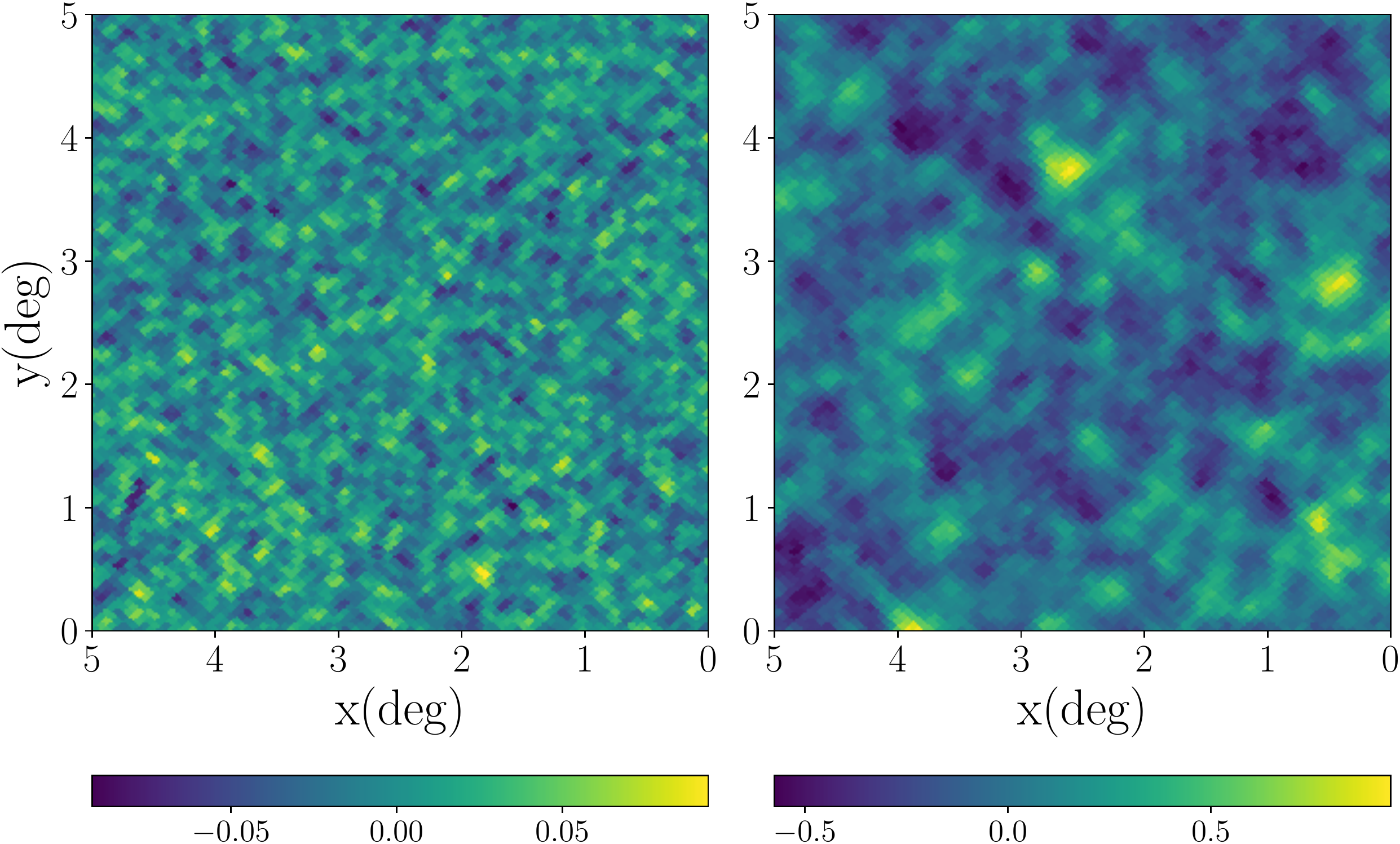}
    \caption{Small 5x5 degree flat sky piece of the curved sky clustering (left) and convergence (right) maps of redshift bin 4 at our fiducial cosmology, using $\mathrm{N_{side}}=1024$ and 5 arcminutes of Gaussian smoothing. The convergence map colourbar shows convergence of each pixel, and the clustering map colourbar shows the galaxy over-density.}
    \label{fig:cartview}
\end{figure}

We replicate DES Y6-like, LSST Y1-like, and LSST Y10-like scenarios in our simulations by applying the corresponding redshift bins, galaxy bias values, and sky fraction. The various scenarios are summarised in Table \ref{tab:smoothing}, and sample simulated convergence and clustering maps are shown in Figure \ref{fig:cartview}.  The ``joint'' case refers to both lensing and clustering.

\subsection{Observables} 

To test the constraining power of MFs, we compare them to the standard $C_{\ell}$ analysis in cosmology. We compare the constraining power of the following three estimators: $C_{\ell}$, MF, and the combination of $C_{\ell}$ and MF. We use the \texttt{NaMaster}\footnote{https://github.com/LSSTDESC/NaMaster} \citep{namaster} library to calculate the full-sky auto-correlation convergence and density angular power spectra $C_{\ell}$ of the fields. We use a bandpower window function with 50 multipoles per bandpower. The maximum $\ell$ value we use is $\frac{3\mathrm{N_{side}}}{2}$, which is 1536 for $\mathrm{N_{side}}$ = 1024, which has a pixel size of 3.4 arcminutes. This means information below 7 arcminutes in the maps is smoothed out.

We give inputs for cosmological parameters, linear galaxy biases, pixel count, fraction of sky, and the amount of Gaussian smoothing applied to the map simulations. 

We convert the integrals from Eq. \eqref{v0}, \eqref{v1}, and \eqref{v2} to the sums in Eq. \eqref{v0_sum}, \eqref{v1_sum}, and \eqref{v2_sum}, calculating the values by taking a sum over the number of pixels $N$:

\begin{equation}
    V_0(\nu_j) = \frac{1}{N} \sum_{i} \Theta(\alpha(\textbf{x}_i)-\nu),
    \label{v0_sum}
\end{equation}

\begin{equation}
    V_1(\nu_j) = \frac{1}{4N} \sum_{i} \Delta(\alpha(\textbf{x}_i)-\nu) \sqrt{\alpha_\phi^2 + \alpha_\theta^2},
    \label{v1_sum}
\end{equation}

\begin{multline}
    V_2(\nu_j) = \frac{1}{2\pi N} \sum_{i} \Delta(\alpha(\textbf{x}_i)-\nu) \\
    \left( \frac{2\alpha_\phi \alpha_\theta \alpha_{\phi\theta}-\alpha_\phi^2 \alpha_{\theta\theta}-\alpha_\theta^2 \alpha_{\phi\phi}}{\alpha_\phi^2 + \alpha_\theta^2} \right),
    \label{v2_sum}
\end{multline}
where $\Delta$ is 1 when $\alpha(\textbf{x}_i)$ is between $\nu_j$ and $\nu_{j+1}$ and 0 otherwise.  The thresholds are evenly spaced from $\mu - 3\sigma$ to $\mu + 3\sigma$, where $\mu$ is the field value mean and $\sigma$ is the field value standard deviation. We use \texttt{HEALPix}\footnote{http://healpix.sourceforge.net} to calculate first and second derivatives of the fields using a series of spherical harmonic transforms \citep{healpy2,healpy1}. 

\subsection{Likelihood} 

\begin{figure}[ht]
    \centering
    \includegraphics[width=\columnwidth]{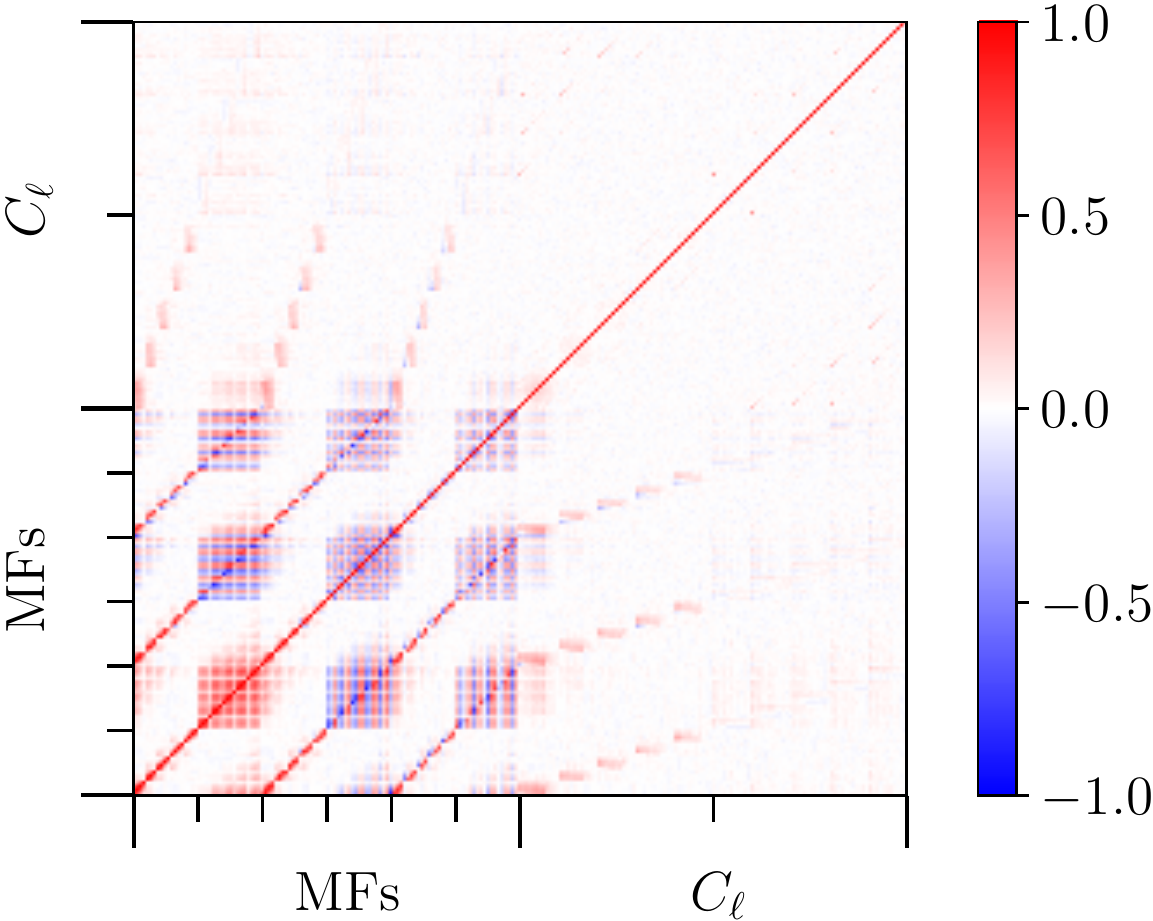}
    \caption{Correlation coefficient matrix of all observables for LSST Y1-like lensing + clustering analysis: 300 MFs and 300 $C_\ell$. There is some correlation between the MFs and some between the MFs and $C_\ell$. The ticks represent the range of the observables for the following statistics: $V_0$ clustering, $V_0$ lensing, $V_1$ clustering, $V_1$ lensing, $V_2$ clustering, $V_2$ lensing, $C_\ell$ clustering, $C_\ell$ lensing. }
    \label{fig:corr}
\end{figure}
We follow a Bayesian procedure to measure our constraining power \citep{bayesian}.  To calculate likelihoods, we measure our observables on a suite of sky map realisations made using DES Y6-like fiducial values for cosmological parameters and biases. Once we have measured the $C_{\ell}$ and/or MFs on each clustering and/or convergence map set, we calculate the likelihood at other cosmologies with a Gaussian likelihood. 

\begin{equation}
    L \propto -0.5 (\vec{x} - \vec{\mu})^t \mathsf{C}^{-1} (\vec{x} - \vec{\mu}),
    \label{eqn:L}
\end{equation} 
where $\vec{\mu}$ represents mean observables over the suite of fiducial realizations, and $\mathsf{C}^{-1}$ is the inverse covariance matrix of the same suite. The simulated data point $\vec{x}$ is evaluated at other non-fiducial values of the input parameters. The calculations of $\mu$ and $\mathsf{C^{-1}}$ vary a random seed in map simulations and require a large number of simulations at the fiducial cosmology. To find the covariance of the statistics, we look at the correlation of the concatenation of the observables: MFs $V_0$, $V_1$, $V_2$, and/or the $C_{\ell}$. The correlation matrix derived from this covariance is shown in Figure \ref{fig:corr}.

The number of fiducial realisations must be greater than or equal to the number of data points for the covariance matrix to be invertible. In our MF analysis, the number of data points is found by multiplying the number of thresholds used by three (the number of functionals) and the number of tomographic bins used, and the length of the $C_{\ell}$ is determined by the map resolution and binning. If there are too few fiducial simulations, the noise in the covariance overwhelms the signal and can cause errors in the constraining power calculation \citep{Petri}. One way to avoid this issue is to use more realisations, but this rapidly becomes time-consuming. 

To make the correction for the finite number of simulations, we follow \citet{Petri} and make the \citet{anderson_2003} adjustment to the inverse: 

\begin{equation}
    \langle \hat{\mathsf{C}}_{*}^{-1}\rangle = \frac{N}{N-p-1} \mathsf{\Sigma}^{-1},
    \label{eqn:cov_fix}
\end{equation} 
where $\Sigma^{-1}$ is the sample inverse covariance matrix, $\hat{\mathsf{C}}_{*}^{-1}$ is an estimator for the inverse covariance matrix, $N$ is the number of realisations minus one (since we find the mean from the data), and $p$ is the number of data points \citep{anderson_2003,covfix}. With $\mathrm{N_{side}}=1024$, there are 300 observables from $C_{\ell}$ and MFs each, so we have 600 data points total. We run the fiducial simulation up to 4000 times for each scenario.

\subsection{Sampling }

We use an MCMC to evaluate posterior values of cosmological parameters based on the output of the Gaussian log-likelihood function in Eq. \eqref{eqn:L}. The cosmological parameter priors can be seen in Table \ref{tab:fiducial} and the bias priors can be seen in Table \ref{tab:redshifts}. For this project we have wrapped the \texttt{emcee} ensemble sampler \citep{emcee} via the cosmological parameter estimator framework \texttt{CosmoSIS} \citep{cosmosis} to run and parallelise the likelihood calculation. We use 40 walkers to generate \texttt{emcee} chains with tens of thousands of samples. As the sampler takes a given number of iterations to explore the parameter space before it settles onto a stationary distribution, we truncate the first several thousand steps of the chains to eliminate the early `burn-in' piece.

Since we are using a simulation to generate the maps and hence observables at each step of our chain, we would ideally marginalise over a large number of possible random seeds for the chain to find the mean observables for a given cosmology or else utilize a more sophisticated simulation-based inference framework (see, e.g. \citealt{cranmer_sbi}). Instead, we use the same random seed for every point in the chain, relying on the fact that with our configuration choices, the \texttt{FLASK} simulated fields are a smooth function of the cosmological parameters. Though this process generates a noisy estimate of the model for a point in parameter space, it does yield a correct unnormalised likelihood, and the sampling acceptance criterion is correct.

Our DES Y6-like model has four redshift bins of source galaxies and five redshift bins of lens galaxies, LSST Y1-like has five redshift bins of source galaxies and five redshift bins of lens galaxies, and LSST Y10-like has five redshift bins of source galaxies and ten redshift bins of lens galaxies. We use only a simple linear model for galaxy bias, with the fiducial bias values calculated using

\begin{equation}
    b_i = \frac{1}{D(\overline{z_i})},
\end{equation}
where $D(\overline{z_i})$ is the growth factor as a function of redshifts at the fiducial cosmology.

\subsection{Masking}

Every survey comes with a sky mask, representing incomplete survey sky coverage, regions blocked by bright stars or other sources, and removals to keep systematic effects out of galaxy maps \citep{des_cosmo}. By limiting the available data, sky masks cause increased variance of fields and affect the estimated MFs values, since MFs are topological descriptors \citep{starmask}. As such, MFs are only effective cosmological probes in simulations if masking is taken into account. 

In this paper we use a simple polar cap sky fraction mask, in contrast to a more realistic analysis with complex small-scale mask features. For speed, we mask pixels after the simulations and derivatives are computed. However, taking derivatives of a field using harmonic transforms is only possible with a complete (unmasked) field.  As such, in an analysis of real data, masked regions must be smoothed and convolved with field values around the mask. This must be carefully addressed in future analyses of real data.

\section{Results}
\label{section:Results}

\begin{figure*}[ht]
\centering
\begin{minipage}{\columnwidth}
    \includegraphics[width=0.95\columnwidth]{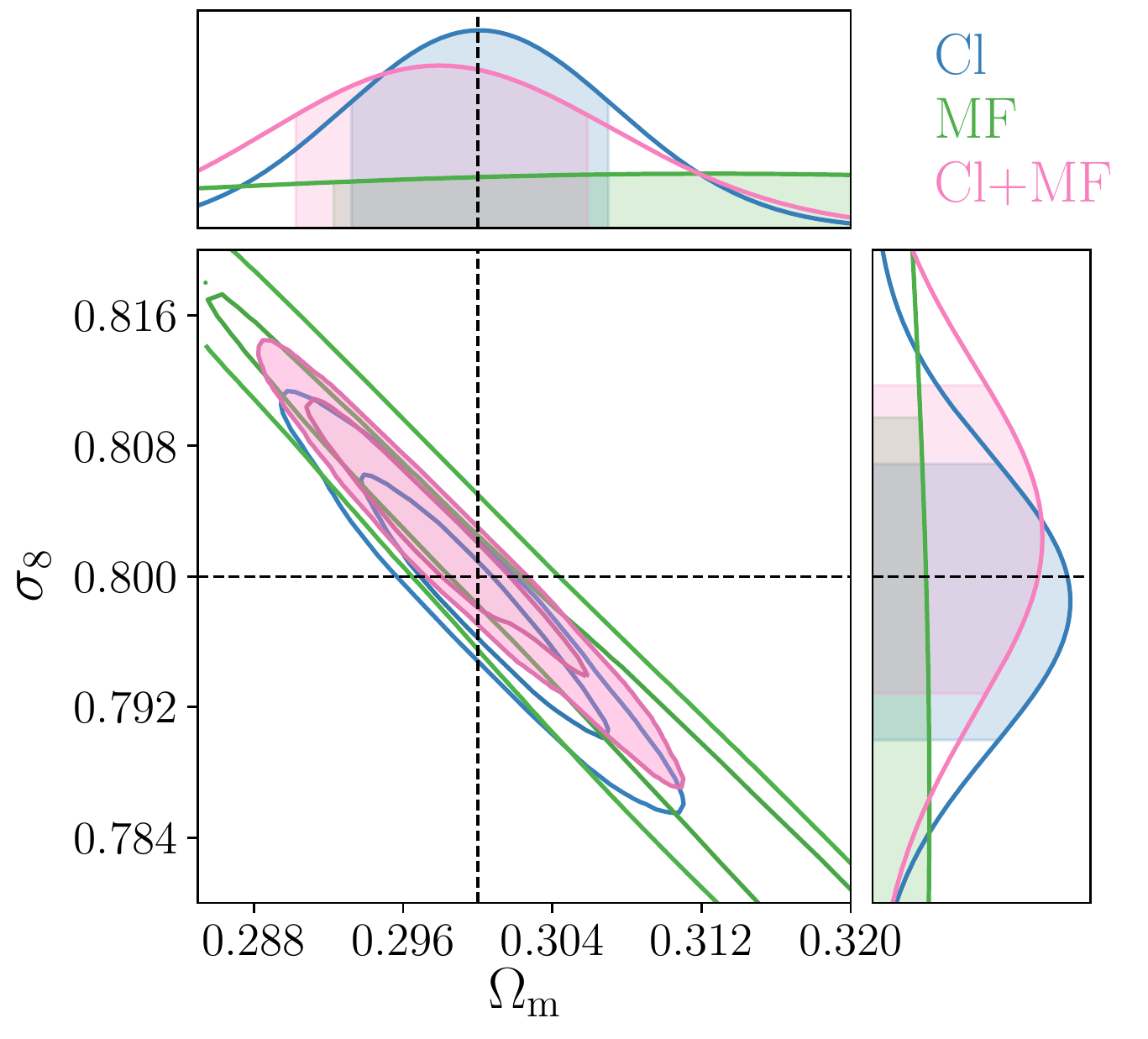}
\end{minipage}
\hfill
\begin{minipage}{\columnwidth}
    \includegraphics[width=\columnwidth]{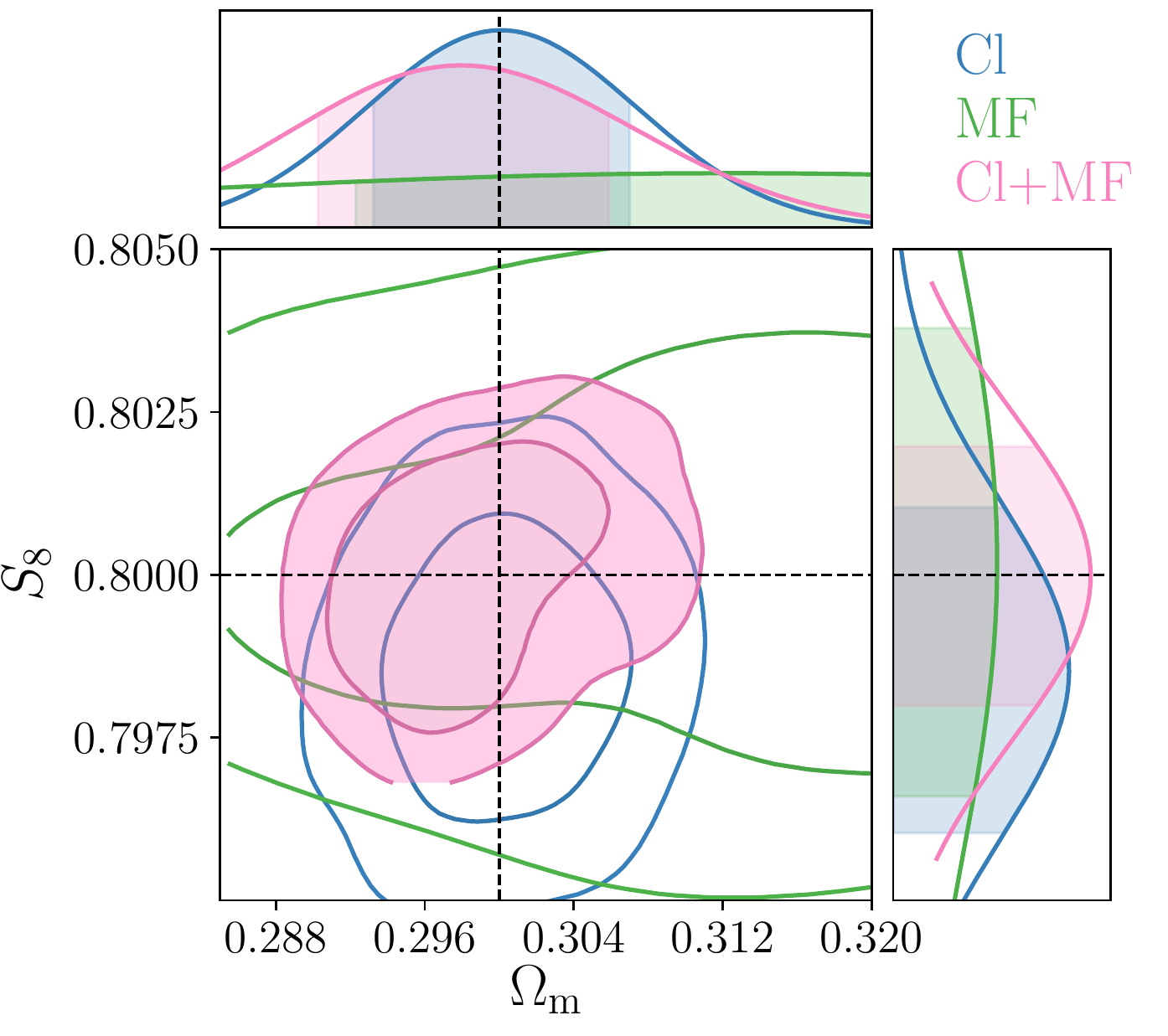}
\end{minipage}
\hfill
\caption{$\Lambda$CDM constraints for power spectra (blue), MFs (green), and joint analysis (pink) measured from convergence + clustering maps using LSST Y1-like redshift bins. Contours show 68\% and 95\% confidence levels. The left plot shows constraints on $\Omega_{\rm m}$ and $\sigma_8$, and the right plot shows constraints on $\Omega_{\rm m}$ and $S_8$. Adding MFs to the 3x2pt analysis does not lead to significantly tighter constraints than $C_{\ell}$ alone, implying MFs and clustering $C_{\ell}$ contain largely the same information.}
\label{fig:contour_a}
\end{figure*}

Figure \ref{fig:contour_a} shows a comparison of constraints on cosmological parameters from three sets of convergence and clustering statistics: MFs alone, $C_{\ell}$ alone, and a combination of MFs and $C_{\ell}$. The $C_{\ell}$ contours are typical for a 3x2pt analysis; MFs alone show the same degeneracy direction at $C_{\ell}$ and show significant constraining power with standard deviations displayed in Table \ref{tab:results}. Adding MFs to the 3x2pt analysis does not lead to significantly tighter constraints than $C_{\ell}$ alone, with only a 15\% decrease in $S_8$ standard deviation, implying MFs and clustering $C_{\ell}$ contain largely the same information. The parameters $\Omega_8$ and $\sigma_8$ separately are relatively poorly constrained.

\begin{figure*}[ht]
\centering
\begin{minipage}{\columnwidth}
    \includegraphics[width=\columnwidth]{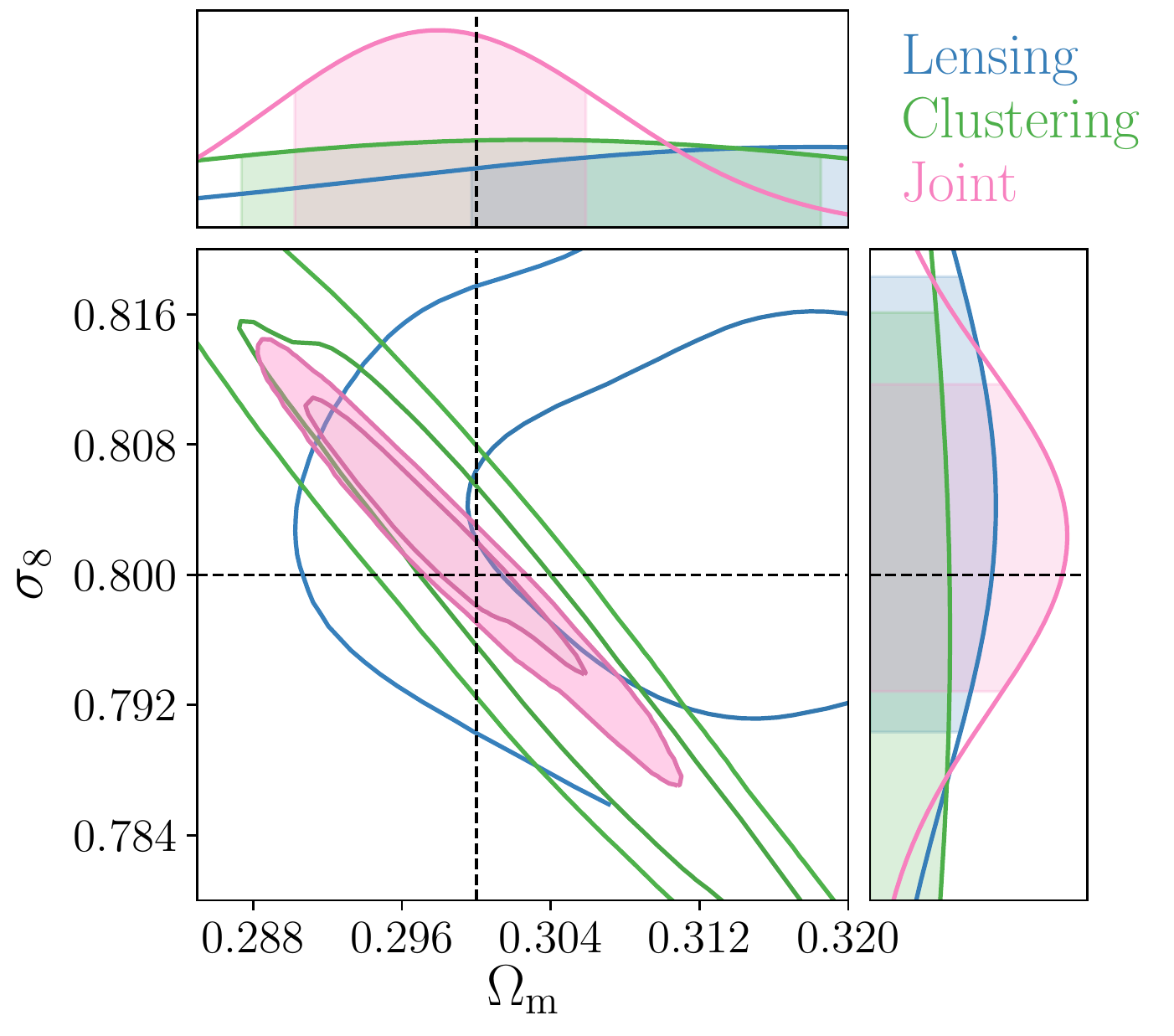}
\end{minipage}
\hfill
\begin{minipage}{\columnwidth}
    \includegraphics[width=0.98\columnwidth]{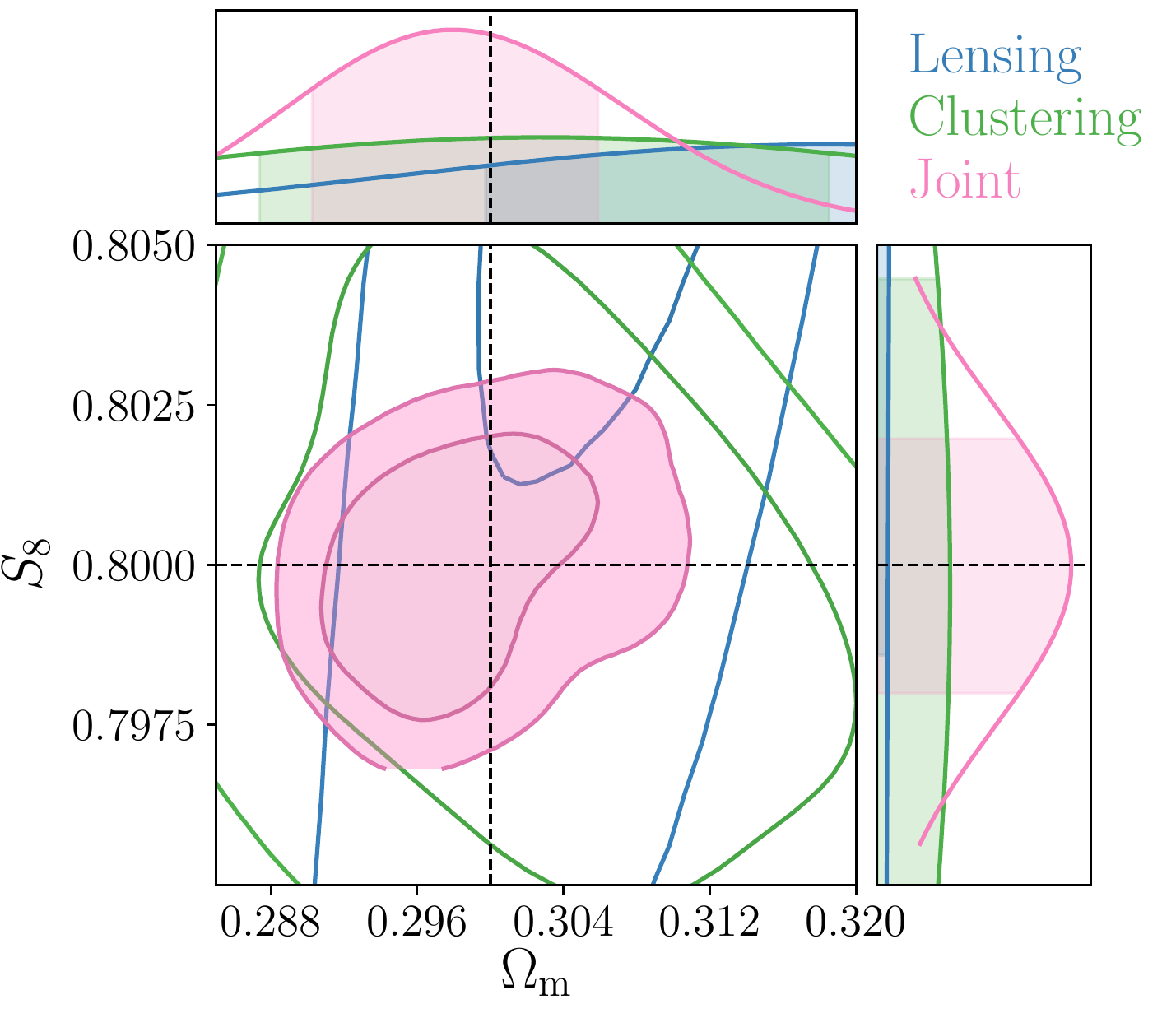}
\end{minipage}
\hfill
\caption{$\Lambda$CDM constraints for joint MF and $C_{\ell}$ statistics measured from convergence (green), clustering (blue), convergence and clustering (shaded in pink) maps using LSST Y1-like redshift bins. Contours show 68\% and 95\% confidence levels. The left plot shows constraints on $\Omega_{\rm m}$ and $\sigma_8$, and the right plot shows constraints on $\Omega_{\rm m}$ and $S_8$. As with standard 3x2pt analyses, adding clustering data to convergence adds significant constraining power to an MF plus $C_{\ell}$ analysis.}
\label{fig:contour_m}
\end{figure*}

Figure \ref{fig:contour_m} compares constraints for the MF and $C_{\ell}$ analysis of convergence, clustering, and a combination of the two. As with standard 3x2pt analyses, adding clustering data to convergence adds significant constraining power to an MF plus $C_{\ell}$ analysis; the joint constraint is stronger than the sum of its parts, pointing to internal degeneracies being broken. The decreased error for the combined model can be compared with higher errors of the individual models in Table \ref{tab:results}, demonstrating again the power of including clustering maps in the simulation.

\begin{figure*}[ht]
\centering
\begin{minipage}{\columnwidth}
    \includegraphics[width=\columnwidth]{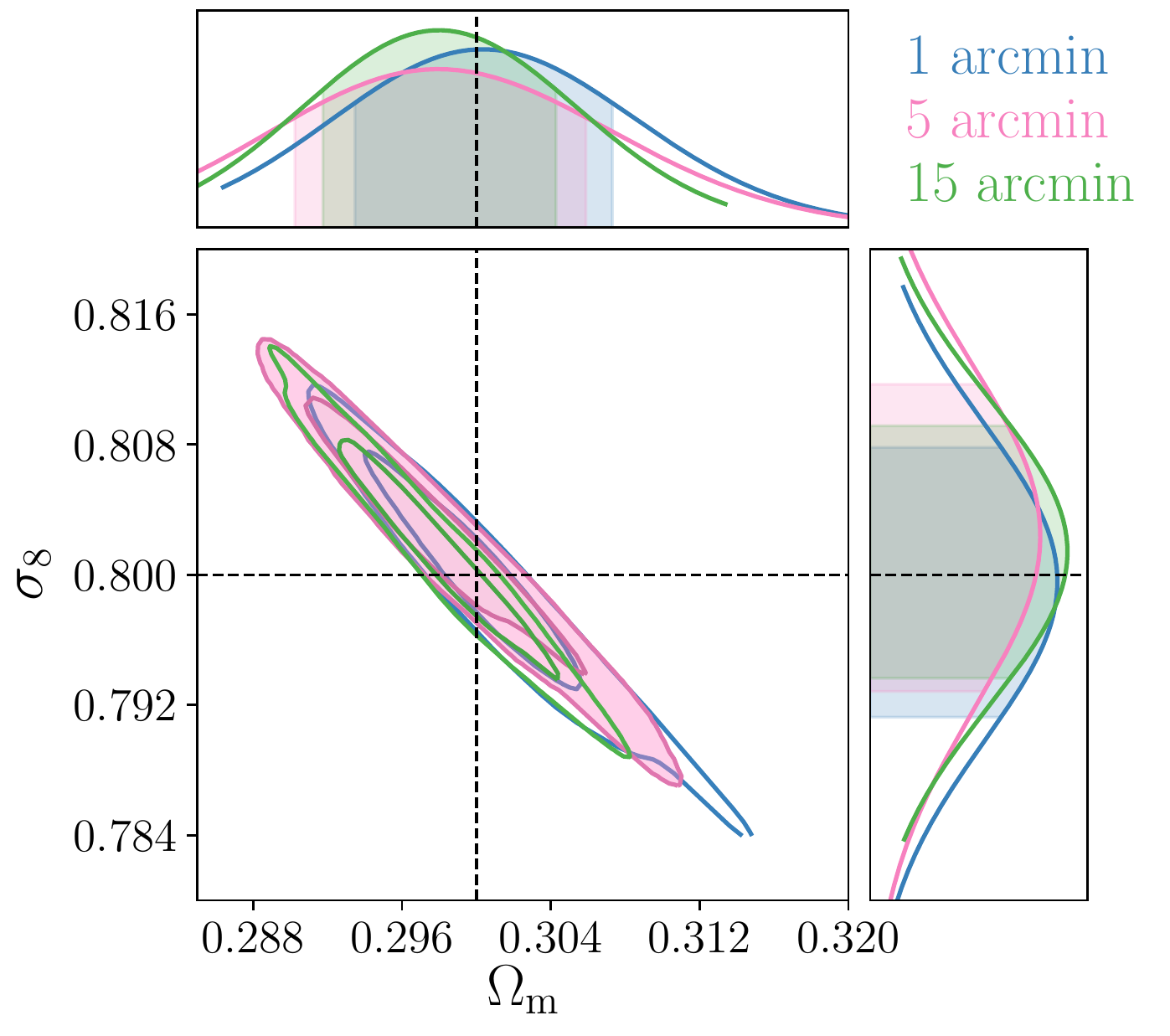}
\end{minipage}
\hfill
\begin{minipage}{\columnwidth}
    \includegraphics[width=\columnwidth]{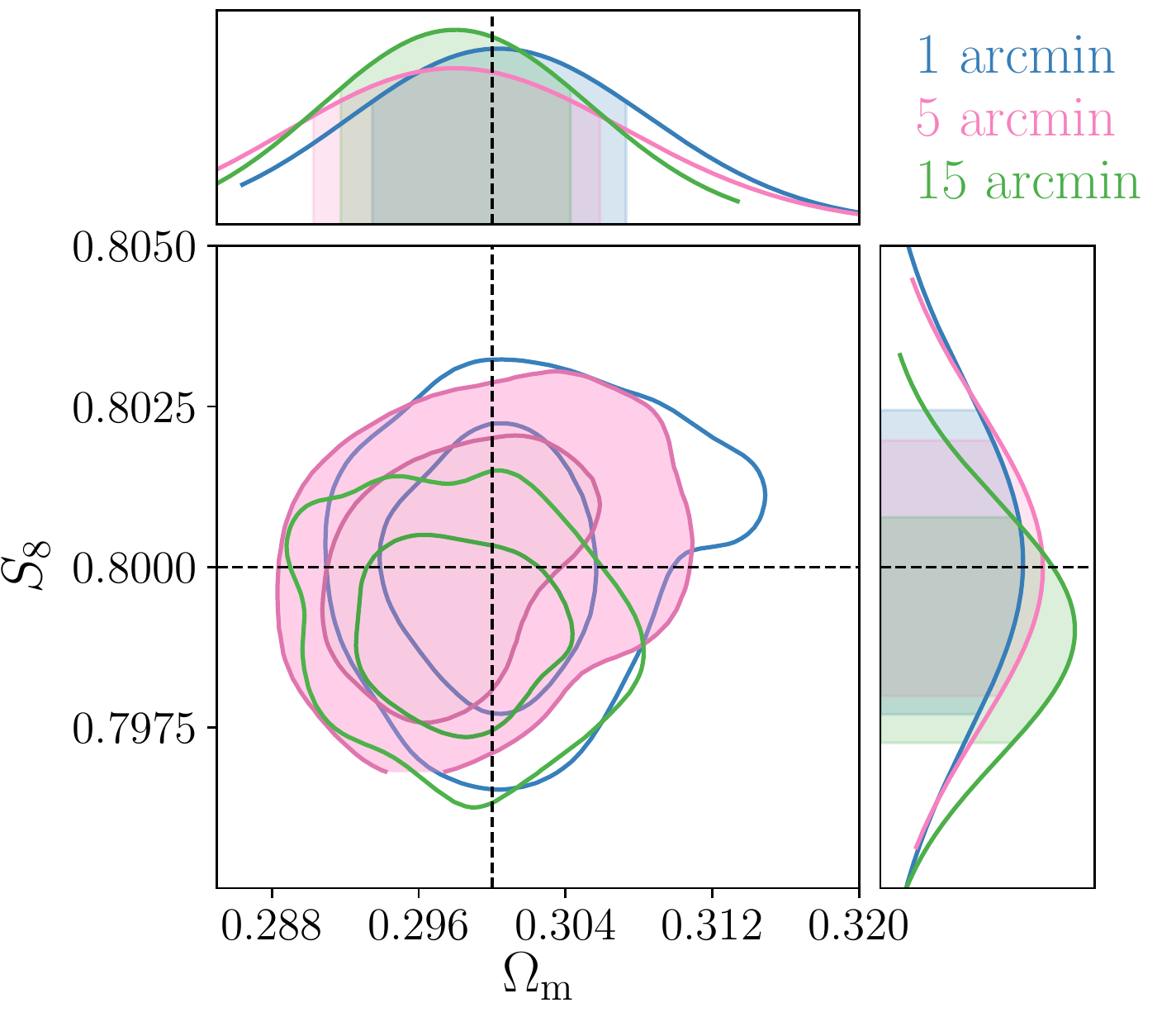}
\end{minipage}
\hfill
\caption{$\Lambda$CDM constraints for joint MF and $C_{\ell}$ statistics measured from LSST Y1-like redshift bins convergence and clustering maps for different Gaussian smoothing levels: 1 arcmin (blue), 5 arcmin (green), and 15 arcmin (pink). Contours show 68\% and 95\% confidence levels. The left plot shows constraints on $\Omega_{\rm m}$ and $\sigma_8$, and the right plot shows constraints on $\Omega_{\rm m}$ and $S_8$. There is not a statistically significant difference in constraining power between the smoothing levels. }
\label{fig:contour_s}
\end{figure*}

Figure \ref{fig:contour_s} shows constraints from the full convergence plus clustering, MF plus $C_{\ell}$ analysis for 1, 5, and 15 arcminute Gaussian smoothing. The resulting uncertainties are quantified in Table \ref{tab:results}, and are not statistically different. This is largely as expected for given our $\mathrm{N_{side}}$ of 1024, which corresponds to 3.4 arcminute pixels, but the noticeable decrease in the $S_8$ constraint size for 15 arcminute smoothing may reflect the different trade-off for smoothing in MF analyses where noise suppression leads to better recovery of the excursion set boundary. With the tomographic number density that we are using here the measurements are noisy even over relatively large smoothing scales - we find the the noise contribution to the value dominates at all the three smoothing scales. Since MFs are higher order statistics, it is not possible to separate out the noise term, but the effect of this noise is included in both the theory and observed values in our MCMC, so does not matter too critically - it is the variation of the noise after that subtraction that affects the constraining power.

The trade-off between larger smoothing kernels to yield lower noise and smaller ones to retain small-scale power depends greatly on the details of the analysis and model.  Understanding this relationship between scale and MF constraining power requires further study.

The constraints for the distributions of DES Y6-like, LSST Y1-like, and LSST Y10-like scenarios are shown in Table \ref{tab:results}. As expected, surveys with more area, bins, and/or depth have stronger constraints.

\begin{table}[ht]
    \centering
    \begin{tabular}{cccc}
    \toprule
    \multicolumn{4}{c}{Errors}\\
    \midrule
         \multicolumn{1}{l}{Analysis Type} & $\Omega_{\rm m}$ &$\sigma_8$& $S_8$ \\
         \midrule
         \multicolumn{1}{l}{Cl} & 0.0040 & 0.0049 & 0.0014 \\ 
         \multicolumn{1}{l}{MF} & 0.0132 & 0.0155 & 0.0019 \\ 
         \multicolumn{1}{l}{Cl+MF} & 0.0045 & 0.0052 & 0.0012 \\ 
         \toprule
         \multicolumn{1}{l}{Map Type} & $\Omega_{\rm m}$ &$\sigma_8$& $S_8$ \\
         \midrule
         \multicolumn{1}{l}{Lensing} & 0.0090 & 0.0118 & 0.0029 \\ 
         \multicolumn{1}{l}{Clustering} & 0.0101 & 0.0075 & 0.0147 \\ 
         \multicolumn{1}{l}{Lensing+Clustering} & 0.0045 & 0.0052 & 0.0012 \\ 
         \toprule
         \multicolumn{1}{l}{Smoothing (arcmin)} & $\Omega_{\rm m}$ &$\sigma_8$& $S_8$ \\
         \midrule
         \multicolumn{1}{l}{1} & 0.0038 & 0.0045 & 0.0012 \\ 
         \multicolumn{1}{l}{5} & 0.0045 & 0.0052 & 0.0012\\ 
         \multicolumn{1}{l}{15} & 0.0034 & 0.0044 & 0.0009\\ 
         \toprule
         \multicolumn{1}{l}{Survey (sky fraction)} & $\Omega_{\rm m}$ &$\sigma_8$& $S_8$ \\
         \midrule
         \multicolumn{1}{l}{DES Y6-like (12.5\%)} & 0.0053 & 0.0072 & 0.0017\\ 
         \multicolumn{1}{l}{LSST Y1-like (44\%)} & 0.0045 & 0.0053 & 0.0012\\ 
         \multicolumn{1}{l}{LSST Y10-like (44\%)} & 0.0006 & 0.0004 & 0.0006\\ 
         \bottomrule
    \end{tabular}
    \caption{\textup{Errors on cosmological parameters for different scenarios.}}
    \label{tab:results}
\end{table}

\section{Conclusion}
\label{section:Conclusion}

In this paper we investigate the impact of including clustering measurements in analyses of Minkowski functionals combined with power spectra on cosmological constraints. Using simulated convergence and clustering maps, we measure the constraining power of the two statistics for DES Y6-like, LSST Y1-like, and LSST Y10-like surveys. While MFs have been previously proven to be useful in convergence map analyses, here we explore their application to photometric clustering maps for the first time. We compare analyses of varying measurement statistics, survey data properties, statistical power, and smoothing levels and present constraints on $\Omega_{\rm m}$, $\sigma_8$, and the better constrained $S_8$. 

The MF measurements have the same $S_8$ degeneracy direction as power spectrum measurements, so we focus on this parameter when inspecting the impact of analysis type, map type, and smoothing amount, as it is generally more robust to analysis choices. We find that MFs probe similar cosmological information to clustering measurements in a 3x2pt analysis, and therefore the improvement on cosmological constraints for $\Omega_{\rm m}$, $\sigma_8$, and $S_8$ is limited, at least in our simplified lognormal map simulation. An important limitation to our analysis, our use of only the auto-correlation $C_\ell$, strengthens this conclusion, since the constraining power of the full 3x2pt analysis is even stronger than that presented here.  Other limitations of our analysis, such as our omission of photo-z and shear-related nuisance parameters, are unlikely to change this conclusion, since they affect both MF and $C_\ell$.

If there is value in including MFs in a 3x2pt analysis, it will perhaps become apparent only when we understand and incorporate small scale effects (i.e. baryons, intrinsic alignments, boost factors, mass reconstruction on small scales, nonlinear galaxy bias) \citep[e.g.][]{osato_2021} combined with noise. That is, MFs are expected to have more potential on such nonlinear scales. 
 
Similar to the standard 3x2pt case, in the $C_{\ell}$+MF analysis we find that the addition of clustering measurements has a significant improvement on the constraints. For the LSST Y1-like case, we find an improvement of over 50\% for all three cosmological parameters. Therefore, we recommend future higher order statistics be measured from both convergence and clustering maps. 

In this project we use curved sky lognormal maps at Gaussian smoothing scales used by \citet{Petri} in convergence mapping, but the analysis has the flexibility to be measured from more sophisticated nonlinear, non-Gaussian data in the future. Our conclusions are significantly dependent on the specific form the non-Gaussian field takes, and so that the realism of the lognormal form of the field matters a great deal. On the intermediate scales that we use, previous research has shown that lognormal fields are a good approximation to real lensing fields \cite{Clerkin_2016}, so the general conclusions drawn here should be reasonable for clustering fields too. 

An MF analysis that included all possible smoothing scales simultaneously would incorporate all the information available in a $C_\ell$ measurement \citealp[see e.g.]{cmb_mf}. While such an analysis is not possible in practice, using a small number of different MF smoothing scales at the same is feasible and could improve the power of MFs.

The maps we use require high computational power, but since the primary constraining power of MFs will come at small scales, using full sky maps may not have been necessary for forecasts. We emphasise, however, that real measurements must consider curved sky effects. 

As such, we make simplifications to the maps for computational efficiency. To usefully apply these statistics to real data one must take into account observing conditions, complex masking, baryon effects, and correlations between redshift bins. This is already true for current surveys like DES, HSC, and KiDS, but will be of heightened importance for upcoming efforts like Rubin, Euclid, and Roman. Masks present a particular challenge for many higher order statistics like MFs, since they increase the complexity of derivatives and other calculations. 

This investigation motivates the use of MFs combined with $C_\ell$ in future analyses of convergence and clustering. Other potential applications for MFs include measuring them from shear maps, instead of convergence maps, or on combinations of fields. For practical data analysis, it is also critical to speed up or avoid the slow likelihood calculations used here, such as by using a faster implementation, an emulator, fitting function, neural network, or use likelihood-free inference. Building on the MF analysis in this work will improve and inform the statistical model used to constrain cosmological parameters when applied to future data.

\vspace{3pt}

\section{Acknowledgments}

NG thanks Alex Hall for the useful discussions and Charlie Mpetha for the motivation.  TT acknowledges support from the Leverhulme Trust. AA received support from a Kavli Fellowship at Cambridge University. We thank the referee for productive discussion.

Results in this paper made use of many software packages, including  \texttt{Numpy}, \texttt{Scipy}, \texttt{FLASK}, \texttt{CCL}, \texttt{CosmoSIS}, \texttt{NaMaster}, \texttt{Emcee}, \texttt{ChainConsumer}, and \texttt{Healpy}/\texttt{HEALPix}.

\bibliography{main}

\end{document}